\documentclass[11pt]{article}

\usepackage{cite}
\usepackage{amssymb}
\usepackage{amsmath}
\usepackage{bm}

\usepackage{graphicx}
\usepackage{epstopdf}
\def\~{\tilde}

\def\be{\begin{equation}}
\def\ee{\end{equation}}

\begin{document}
\title{The Metaplectic Group, the Symplectic Spinor and the Gouy Phase.}
\author{M.\ Fernandes,\vspace{0.1cm}\\ Instituto de F'sica, Universidade de Bras'lia, \\70910-900, Bras'lia, DF, Brazil\vspace{0.2cm}
\\and\vspace{0.2cm}\\ B. J. Hiley\vspace{0.1cm}
\\TPRU, Birkbeck College, University of London\\ Malet
Street, London WC1E 7HX}
\date{}

\maketitle

\begin{abstract}

	In this paper we discuss a simplified approach to the symplectic Clifford algebra, 
the symplectic Clifford group and the symplectic spinor by first  extending the Heisenberg algebra.  We do this by adding a new idempotent element to the algebra.  It turns out that this element is the projection operator onto the Dirac standard ket. When this algebra is transformed into the Fock algebra, the corresponding idempotent is the projector onto the vacuum. These additional
elements give a very simple way to write down expressions for symplectic spinors as elements of the symplectic Clifford algebra.  When this algebra structure is applied to an optical system we find a new way  to understand the Gouy effect.  It is seen to arise  from the covering group, which in this case is the symplectic Clifford group, also known as the metaplectic group.  We relate our approach to that of Simon and Mukunda who have shown the relation of the Gouy phase to the Berry phase.  

\end{abstract}

\section {Introduction.}

In this paper we want to use the relatively unknown symplectic Clifford algebra \cite{ac90} to address the Gouy effect \cite{ gg90} \cite{rsnm93}. This effect is a consequence of the discontinuity of the phase as one passes through a focal point of an optical system.  It arises from the double cover of the symplectic group, namely, the metaplectic group.  This problem can be handled with standard group theory techniques but we want to use the symplectic Clifford algebra [SCA] to bring out the analogy with the way the double cover of the orthogonal group is treated with the much more familiar orthogonal Clifford algebra [OCA].  These algebras are well known since they are used to describe spin through the Pauli matrices and the Dirac electron using the gamma matrices.

The mathematical structure of the OCA and its uses in physics is well documented in the literature (see, for example, Lounesto \cite{pl01}, Porteous \cite{ip95} and Crumeyrolle \cite{ac90}).  One of the interesting
aspects of this algebra is the appearance of the Clifford  group, known also as the spin group.  It is this group that provides the double cover of the corresponding orthogonal group, which in turn provides the properties of spinors that play such a vital role in the physics of fermions.  It also produces an effect that is analogous with the Gouy effect, namely, 4$\pi$ periodicity of the spinor which has been so beautifully demonstrated in the experiment of  Rauch {\em et al} \cite{rwbf}.
  
	The SCA, on the other hand, has received relatively little attention in the physics literature.  It has, however, received detailed treatment in the mathematical literature, but these discussions  depend on techniques that are not that familiar to physicists and are, in
consequence, difficult to relate to actual physical situations.  Perhaps the most accessible details can be found in Crumeyrolle \cite{ac90}.

	In this paper we wish to present a simplified approach to the SCA which will enable it to be used in physics in a straight forward manner.  In particular we will show how the metaplectic group arises as the symplectic Clifford group, providing a double cover of the symplectic group (See also Bacry
and Cadilhac \cite{bc81}).  This gives rise to the corresponding symplectic spinor which, in turn, provides a basis for discussing the Gouy effect  \cite{ gg90}.  This effect has recently been linked with the Berry phase by
Simon and Mukunda \cite{rsnm93}.  We will show how this phase change arises from the properties of the symplectic spinor.

Although the symplectic spinor itself has been discussed in a
mathematically rigorous way by Kostant \cite{bk74}, Guillemin \cite{vg78} Guillemin and Sternberg \cite{vgss84},  Crumeyrolle \cite{ac90}, and Robinson \cite{pr92}, we will present some of these ideas in a much simpler form which will make the algebraic approach more accessible to physicists.  This will enable us to see much more clearly the role played by the symplectic spinor in physical situations.  To do this we will relate our algebraic methods to the more familiar approach using the Dirac bra-ket notation,
thus clarifying the relationship of these methods to those using the SCA directly.
  
	The method that we will use here has already been outlined in Frescura and Hiley \cite{ffbh84}, but we are now able to present these ideas in a simpler way that can be directly related to the approach used by Simon and Mukunda  \cite{rsnm93}.  The method itself has been strongly influenced the
papers of Sch\"{o}nberg \cite{ms57}.  He was more interested in developing a quantum geometry, but we will not discuss this aspect here.  Rather we want to establish the relation between the algebraic approach and the more usual approach used in physics.
  
	In section 2 we define the symplectic Clifford algebra, while in section 3 we discuss  the symplectic Clifford group and its precise relation to the metaplectic group.  Here we present the generators of the symplectic group and the corresponding generators of the metaplectic group.  In section 4 we define the algebraic symplectic spinor in terms of an extended Heisenberg algebra.  This extension requires the introduction of a primitive
idempotent, the meaning of which emerges through a consideration of a discrete Weyl algebra. 

This algebraic spinor is then related to physicist's approach using the bra-ket notation in section 5.  We also show that the primitive idempotent that we introduce is the projector onto the Dirac standard ket [16].  In section 6 we demonstrate the physical consequences of the double cover, while in section 8 we relate the algebraic approach to wave theory thus giving a complete understanding of the relation between the algebraic approach and the physics of the Gouy effect.  Finally we relate our work to the discussion of the Berry phase given in Simon and Mukunda  \cite{rsnm93}.

\section {The Geometric or Clifford Algebras}

\subsection{Two Geometric Algebras.}

We start with a simple remark.  Consider a bilinear form, $B$, defined by
\begin{eqnarray*}
B\;:(x,y)\in E\times E\rightarrow B(x,y)\in{\bm K}.
\end{eqnarray*}
where $E$ is an $n$-dimensional vector space and ${\bm K}$ is either $\mathbb C$ or $\mathbb R$.  Then it can be proved \cite{ac90} that only two geometries arise which are distinguished by whether $B$ is symmetric or antisymmetric.  If $B$ is symmetric then the geometry is called orthogonal; if $B$ is antisymmetric, the geometry is called symplectic.  These geometries arise in both classical and quantum physics.  

In classical physics we find that, in general, all the relevant dynamical equations can be obtained from the group structure, but in quantum physics we must consider the covering group.  It is through the covering groups that all the important gauge effects emerge.  Furthermore in quantum physics, as pointed out by Weyl \cite{hw31},  it is the group algebra that plays a key role.  This is essentially because of the importance of linear superposition in quantum mechanics.

To discuss the mathematical properties of a group algebra, we go to the geometric algebra, to the Clifford algebra.  Since there are two basic geometries, we should not surprised to find two types of Clifford algebras, the symplectic and the orthogonal.  Historically the importance of the orthogonal Clifford algebra in physics slowly emerged, first with the introduction of the Pauli spin algebra, then in the description of the Dirac electron and   more recently, in the Penrose twistor structure \cite{rp72}.

What is not so generally well known, except to those working in the field of quantum optics, is the role played by the symplectic Clifford algebra.  In this paper we want to concentrate on this algebra, bringing out the role played by the metaplectic group and the symplectic spinor.  The aim being to bring out the more general structure which has been seen to only to be useful in quantum optics.

\subsection{The Orthogonal Clifford Algebra.}

Let us start with the more familiar orthogonal Clifford algebra, ${\cal C}(g)$, that plays a central role in the theory of  the Dirac electron.  We start from an $n$-dimensional vector space $E$ over the field $K$ with a diagonal metric tensor $g_{ij}$ comprising of $\pm 1$ terms down the diagonal.  We take an orthonormal basis, $x_i$ of this space and map it onto an algebraic structure $\alpha :x_i\rightarrow e_i$ subject to the multiplication rule
\begin{eqnarray}
e_i \circ e_j+e_j\circ e_i=g_{ij} 		\label{qe:anticom}
\end{eqnarray}
Here $\circ$ is called the (orthogonal) Clifford product.
The well known examples are the $e_i$ are the Pauli spin matrices in the non-relativistic case with the metric $(+1,+1,+1)$ and the Dirac gamma matrices in the relativistic space where the metric is the usual $(+1,-1,-1,-1)$.

We are interested in forming tensors, $\otimes E$ and to handle these objects we form
\begin{eqnarray}
\alpha :\otimes E\rightarrow \otimes E/{\cal N}(Q) ={\cal C}(Q)	\label{eq:defC}
\end{eqnarray}
where ${\cal N(Q)}$ is the two sided ideal generated by
\begin{eqnarray*}
x\otimes x-Q(x),\quad x\in E\subset \otimes E
\end{eqnarray*}
Here $Q(x)$ is the quadratic form that defines $g_{ij}$.  

This structure enables us to handle the antisymmetric tensors formed on $\otimes E$.  A general element of the algebra can be written in the form
\begin{eqnarray*}
\Gamma= G_0+g_ie_i + G_{[ij]}e_ie_j +\dots + G_{[ij\dots n]}e_ie_j\dots e_n,
\end{eqnarray*}
where $G_{[ij\dots]}$ is an antisymmetric tensor of finite rank.

\subsection{The Symplectic Clifford Algebra} 

In order to encompass the symmetric tensors, let us start with the vector space $ W $ which we assume can be written in the form $W=U\oplus V$, where $U$ and $V$ are vector space, each of dimension $n$.  In order to construct an algebra based on this vector space, let us introduce a mapping $\sigma :W\rightarrow ({\cal A}, \bullet)$ such that
\begin{eqnarray*}
\sigma(x) \bullet \sigma(y)  -  \sigma(y) \bullet \sigma(x)  =  h\omega(x,y) 
\end{eqnarray*}
where $x$ is a basis in $U$ and $y$ is a basis in $V$, while $\omega$ is the two form constructed in the following way,
\begin{eqnarray*}
\omega(w_1,w_2) = x_1\cdot y_2-x_2\cdot y_1.
\end{eqnarray*}
Here $\bullet$ is the symplectic Clifford product and $\cdot $ is the usual scalar product in a vector space.

We can now introduce a basis such that 
\begin{eqnarray*}
\sigma :    x_{j} \rightarrow  e_{j} \hspace{1.5cm} \mbox{and} \hspace{1,5cm}  \sigma : y_{j} \rightarrow  e_{j}^{*}
\end{eqnarray*}
Then
\begin{equation}
	[ e_{i}  ,  e_{j} ]_-  =  0, \hspace{1.5cm}          [ e_{i}  , e_{j}^{*}]_-  = 
\delta_{ij}, \hspace{1.5cm}        [e_i^{*} , e_{j}^{*}]_-  = 0.	\label{eq:com}
\end{equation}

Here $ \omega( e_{i} , e_{j}^{*})  =  \delta_{ij} $.    The set of elements, $\{e_{i} ,
e_{i}^{*}\} $ can be regarded as the generating elements of the symplectic Clifford algebra, $ {\cal C}_s $.  By a change of notation, we can put this algebra in a much more familiar form by writing
\begin{eqnarray}
\sigma(x_j)=Q_j \quad\mbox{and}\quad\sigma(y_j)=iP_j/\hbar	\label{eq:QP}
\end{eqnarray}
 we recognise the commutation relations define nothing more than the Heisenberg algebra.  In other words what we regard as something which is the essence of quantum phenomena is nothing but the symplectic structure arising from the group algebra of the symplectic geometry.

To link up with the formal definition of the symplectic Clifford algebra \cite{ac90}, we consider the tensor algebra  $ \otimes $ $ W $ by the two sided ideal $ \Im $  generated by the elements
\[	x \otimes  y  - y  \otimes  x  - h  \omega  (x,y)  \hspace{2cm} 			x , y  \in  
W.\]
write
\begin{equation}
\sigma  :  \otimes W \rightarrow \otimes W/\Im = {\cal C}_s
\end{equation}

A general element of the algebra can then be written as
\[ \Lambda =  \sum_{h_{i}=1}^{\infty} \sum_{k_{i}=1}^{\infty} \lambda_{h_{1} , h_{2}...h_{r} ,
k_{1}, k_{2}....k_{r}} (e_{1})^{h_{1}}(e_{2})^{h_{2}}.......(e_{r})^{h_{r}}(
e_{1}^{*})^{k_{1}}( e_{2}^{*})^{k_{2}}....( e_{r}^{*})^{k_{r}}, \]
which for convenience, we will write as
\be
	\Lambda  =   \sum \lambda_{HK^{*}} e^{H} e^{K^{*}}.				
\ee
Since this is an infinite series,  one must be careful and work only with elements that are bounded.  At this stage we will simply assume that all relevant functions are elements that rapidly decrease as $ h_{i} \rightarrow \infty $  and $ k_{i} \rightarrow \infty.$  We will not discuss these technical problems further as they are treated in detail in Crumeyrolle \cite{ac90}. 
Without this condition we have a structure which is called a quasi-algebra. 

By now the similarities and differences between the two Clifford algebras should be clear.  What should also be clear is that the two algebras are complementary, one dealing with the antisymmetric tensors, while the other treats the symmetric tensors.  To do this the orthogonal Clifford algebra exploits the anti-commutator brackets, while the symplectic Clifford exploits the commutator brackets, suggesting there is a deep structural connection between quantum processes and the underlying geometry, a feature which is hidden in the conventional Hilbert space formalism.  This feature was first brought to our attention by the extensive work of Sch\"{o}berg \cite{ms57}. 

 In this paper we will not be concerned with these much deeper questions.  Rather we will study the relatively unknown SCA, bringing out those features that have parallels in the more familiar orthogonal Clifford algebra.
 One of the difficulties of dealing with the SCA as opposed to the OCA is that it is infinite dimensional whereas the OCA is finite dimensional which makes it much easier to discuss.  The question then arises as to why we should bother with this complicated structure.  
 
 The answer is extremely simple.  The symplectic symmetry lies at the heart of the kinematics of both classical and quantum physics.  In focusing almost exclusively on the Schr\"{o}dinger equation we loose any connection with classical Hamilton-Jacobi theory.  Yet it is possible to lift the classical symplectomorphism onto the covering structure and find the Schr\"{o}dinger equation arising as a natural group of automorphism in this covering structure.  This remark is motivated by the extensive mathematical work of Guillemin and Sternberg \cite{gs77} \cite{vgss84}, de Gosson \cite{mdg02}  and de Gosson and Hiley \cite{mgbh11}.  It is with this background in mind that we will now discuss the SCA in some detail.

\section{ The Symplectic Clifford Group, the Metaplectic Group}

\subsection{The Orthogonal Clifford Group.}

What we will concentrate on in this paper is the properties of the double cover of the symplectic group.  We have already remarked that a covering group arises naturally in the corresponding group algebra, the symplectic Clifford group. To see how this all works, we will first briefly discuss the more familiar OCA. 

We start by defining the orthogonal Clifford group (OCG) to be the multiplicative group of invertible elements $g\in {\cal C}(Q)$ such that 
\begin{eqnarray}
\phi_g(x)=\alpha(g)xg^{-1}\in E,\quad\forall x \in E		\label{eq:OCG}
\end{eqnarray}
where $\alpha$ is the main anti-involution\footnote{A formal definition of the main anti-involution of ${\cal C}(Q)$ will be found in Porteous \cite{ip95} }
of ${\cal C}(Q)$. 
This group is related to, but bigger than, the spin group that we use in physics.  However if we define the spin norm by $N(g)=\alpha(g)g$, then the spin group is the subgroup of the Clifford for which $N(g)=+1$. We find the spin group is generated by the bivectors of the algebra.

Now as already pointed out in the introduction, it is the spin group which explains the $2\pi - 4\pi$ periodicity of the spinor that has been demonstrated in the neutron interference experiments of Rauch {\em et al} \cite{rwbf}.  To see how this is explained through the OCA, we need to introduce the concept of the algebraic spinor (For a more general discussion see Frescura and Hiley \cite{ffbh80} and Benn and Tucker \cite{ibrt83}). 

 First note that in order to use equation (\ref{eq:OCG}), we need to express the state of the particle by an element of the algebra.  As explained in Hiley and Callaghan \cite{bhrc11} and in Hiley \cite{bh13}, we replace the usual Pauli spinor by an element of a minimal left ideal defined by
\begin{eqnarray*}
\Phi_L=\phi_L\epsilon=(g_0+g_1e_{23}+g_2e_{13}+g_3e_{12})\epsilon
\end{eqnarray*}
where $\epsilon$ is a primitive idempotent which we will choose to be  $\epsilon =(1+e_3)/2$ thus defining a direction in space.   The $e_i$ are the generators of the Pauli Clifford algebra, while  the  $g_i$ are related to the two components of the usual Pauli spinor by 
\begin{eqnarray*}
 g_0=(\psi_1^*+\psi_1)/2\quad\quad g_1=i(\psi_2^*-\psi_2)/2\\
 g_2=(\psi_2^*+\psi_2)/2\quad\quad g_3=i(\psi_1^*-\psi_1)/2
\end{eqnarray*}
Now a general element of the OCA can be written in the form
\be
A=\Phi_L\Phi_R=\phi_L\epsilon\phi_R	\label{eq:8}
\ee
A  rotation of $A$ through an angle $\theta$ is given by
\be
A'=\alpha(g(\theta/2))Ag(\theta/2)=[\alpha(g(\theta/2))\phi_L(x)\epsilon][\epsilon\phi_R(x)g(\theta/2)]
\ee
showing that the left ideal, the spinor is rotated by $\alpha(g(\theta/2))$, while the right ideal, the conjugate spinor is rotated by $g(\theta/2)$.  Thus when we  rotate through an angle $\theta=2\pi$, we find that $\Psi_L$ changes sign, but when $\theta=4\pi$,  the sign change disappears showing we have returned to the original state.  

\subsection{The Symplectic Clifford Group.}

If we want to find the physical consequences of the double cover of the symplectic Clifford group [SCG], we start by recalling the definition of a symplectic transformation $ S \in Sp(2n) $  as 
\begin{eqnarray*}
\omega(S(w_{1}) , S(w_{2}) )  =  \omega(w_{1}, w_{2} ),
\end{eqnarray*}
where $ \omega(w_{1}, w_{2})$ is the skew-symmetric form defined above.

Following the analogy with the orthogonal Clifford group, we now define an invertible element, $\mu_s\in {\cal A}_s$, of the SCG, $ G_{s} $, 
\begin{eqnarray*}
\mu_{s}w\mu_{s}^{-1} \in W, \hspace{1.5cm}    	\forall \hspace{0.5cm}  w \in W
\end{eqnarray*}
These elements provide a homomorphism from $ G_{s} $ into Sp(n), the symplectic group via
\begin{eqnarray*}
\rho_{s} : \mu_{s} \rightarrow \rho_{s}(\mu_{s})\hspace{1cm} 	{\rm with} \hspace{1cm} 
\mu_{s}w\mu_s^{-1}  =  \rho_{s}(\mu)(w)   =  S(w).
\end{eqnarray*}
 
 If we restrict ourselves to those $ \mu_{s}   =   M_{s} $  whose norm is unity, we have
the elements that form the metaplectic group Mp(n) \cite{rsnm93}.  This group is the symplectic analogue of the orthogonal spin group.
Thus a metaplectic transformation can be written as
\begin{eqnarray*}
\bm M_{s} w \bm M_{s}^{-1}  =  \bm Sw \hspace{2.5cm}			w  \in W.
\end{eqnarray*}
Using the basis elements $ e_{i}=Q_i $ and $ e_{i}^{*}=P/\hbar $, equation (\ref{eq:QP}) can be written in a more transparent form.  To do this let us form
\begin{eqnarray*}
w  = \left(\begin{array}{crcr} u \\ v \end{array}\right)  = 
 \left(\begin{array}{crcr}  Q \\  P \end{array}\right)
\end{eqnarray*}

	As is the case of the spin OCG, the metaplectic group is 
generated by quadratic elements of the algebra \cite{bk74}.  For the special case of Mp(2), the generators are
\begin{eqnarray*}
H  =  \frac{1}{2} [ P Q  +  Q P] ; \hspace{1.52cm}	    X  = 
\frac{ Q^{2}}{2}  ; \hspace{1.5cm}	 Y  =  \frac{ P^{2}}{2}
\end{eqnarray*}
so that
\begin{eqnarray*}
	[X, Y]  =  H;\quad\quad [H, X]  =  2X;\quad\quad[H, Y]  =  - 2Y
\end{eqnarray*}
which will immediately be recognised as the Lie algebra of the symplectic group, Sp(2).

Let us now collect together the relevant formulae for the general case in $2r$ dimensions.  Consider a matrix $ \bm S\in Sp(2r)$.  This can be written in the form
\be
 \bm S =  \left( \begin{array}{clcr} {\bm A} &{\bm B} \\ {\bm C} & {\bm D} 
\end{array}\right) 
\ee
where ${\bm A},{\bm B},{\bm C},{\bm D}$ are $ (r \times r) $ matrices.  For ${\bm S}$ to be an
element of the symplectic group, the following conditions must be satisfied:
\[{\bm A}{\bm D}^{t}   -  {\bm B}{\bf C}^{t}   = {\bm I} \]
\[{\bm A}^{t}{\bm C}   =  {\bm C}^{t}{\bm A} \hspace{2cm}		{\bm A}{\bm B}^{t}  =  {\bm B}{\bm A}^{t} \]
\[{\bm B}^{t}{\bm D}   =  {\bm D}^{t}{\bm B} \hspace{2cm}		{\bm	C}{\bm D}^{t}  =  {\bm
D}{\bm C}^{t} \]
where $ {\bm X}^{t} $ means transposition.
It is convenient to write a general symplectic matrix in the canonical form
${\bm S}  =  {\bm S}_{1}{\bm S}_{2}{\bm S}_{3} $
where
\be
	{\bm S}_{1}  =  \left(\begin{array}{cc}{\bm I} & {\bm B\bm D}^{-1}\\{\bm 0} & {\bm I}\end{array}\right)\hspace{1cm}    {\bm S}_{2}  = \left( \begin{array}{cc} 
{\bm D}^{-1} & {\bm 0} \\{\bm 0} & {\bm D}\end{array}\right) \hspace{1cm} {\bm S}_{3}  =  
\left(\begin{array}{cc} {\bm I} & {\bm 0}\\ {\bm C\bm D}^{-1} & {\bm I}\end{array}\right)
\ee	
		
It is not difficult to show that the corresponding elements of the metaplectic group are
\be
{\bm M}_{{\bm s}_{1}}  =  \exp[-\frac{i}{2}\left(\bm B\bm D^{-1}
\right)^{jk} P_{j} P_{k}]	\hspace{1.6cm}	 	 \leftrightarrow
\hspace{1cm}\left(\begin{array}{clcr}{\bm I} & {\bm B\bm D}^{-1}\\ {\bm 0} & {\bm
I}\end{array}\right)\ee
\be
{\bm M}_{{\bm s}_{2}} = \exp[\frac{i}{2}\left({\bm D}^{jk}\right)(P_{k}
Q_{j}  +  Q_{j} P_{k})] \hspace{0.7cm} 	\leftrightarrow\hspace{1cm}
\left(\begin{array}{cc} {{\bm D}^{-1}} & {\bm 0}\\ {\bm 0} & {\bm D}\end{array}\right)
\ee
\be
	{\bm M}_{{\bm s}_{3}}  =  \exp[\frac{i}{2}\left(\bm C\bm D^{-1}\right)^{jk}
Q_{j} Q_{k}]\hspace{1.7cm}  \leftrightarrow \hspace{1cm}		\left(\begin{array}{cc} {\bf I} &
{\bm 0}\\ {\bm CD}^{-1} & {\bm I}\end{array}\right)
\ee
This then is the canonical decomposition of the metaplectic group which satisfies the relation
\be
 {\bm M}_{\bm s} \left(\begin{array}{cc} Q\\  P\end{array}\right){\bm M}_{\bm s}^{-1}  = 
\bm S  \left(\begin{array}{cc} Q \\  P \end{array}\right)
\ee
The metaplectic group gives rise to a projective representation,
\be
{\bm M}_{{\bm s}\prime}{\bm M}_{s}  =  \lambda({\bm S}',{\bm S}) {\bm M}_{{\bm s}'{\bm s}}
\ee			
where $ \lambda $ is a complex number of modulus unity.  It can be shown that $ \lambda $ 
takes two  values $\pm 1$, so that

	\[ {\bm S} \rightarrow\left \{  \begin{array}{cc} {\bm +M_{\bm s}}\\ {\bm - M_{\bm s}} \end{array}\right. \]
giving us a 2-valued representation showing that  the metaplectic group is the double cover of the symplectic group.

\section{The Symplectic Spinor}

	We now want to construct an algebraic symplectic spinor in direct analogy with the more 
familiar algebraic orthogonal spinor. 
There is one difficulty that we immediately encounter in trying to follow this proceedure. 
The OCA is a non-nilpotent algebra and therefore there exists at least one primitive
idempotent within the algebra.  This idempotent enables a minimum left ideal to be defined and
it is this minimum left ideal that gives rise to the algebraic spinor.
  
	The SCA as defined is a nilpotent algebra and in consequence it contains no primitive 
idempotent (see Frescura and Hiley \cite{ffbh80a}).  Therefore it is not possible to construct an algebraic symplectic spinor without extending the algebra. In order to see how this can be doner, let us first consider the maximal left ideal defined by

	\[ \Im_{m}  =   \{ \Lambda \in C_{s},\hspace{0.2cm}  \Lambda = \sum \lambda_{HK^{\star}} 
e^{H}  e^{K^{\star}},\hspace{0.2cm}  {K^{\star}} \neq 0\}\]
The quotient space $C_{s}/\Im_{m} $ forms a spinor representation which is irreducible because 
$ \Im_{m} $
is maximal. 

 In order to develop the formalism in analogy to the OCA, Crumeyrolle \cite{ac90a} writes
$ C_{s}/\Im_{m} $  as $ C_{s}{\Phi^{\star}} $ where $ {\Phi^{\star}} $ is regarded as a symbol
resembling the isotropic $r$-vector of the orthogonal case.  The space $ C_{s}{\Phi^{\star}} $ is
linearly isomorphic to the symmetric algebra on the vectors $ ( e_{1},   e_{2},
...... e_{r} ).$  Thus a symplectic spinor can be written in the form
\be
		\psi  =  \sum \lambda_{H} e_{H}{\Phi^{\star}}		\label{eq:SMI}		
\ee

	As we have already remarked, our own approach was motivated by the original proposal  of Sch\"{o}nberg \cite{ms57}.  He has suggested that we  extended the SCA as defined above by introducing an element $E$, satisfying  the conditions
\[	E^{2}  = E;  \hspace{1cm} \sigma (y)E  =  0  \hspace{1cm} {\rm and} \hspace{1cm}
 	E \sigma(x)  =  0.		\]
In terms of the variables defined in equation (\ref{eq:QP}), viz
	\[ \sigma(x_{i})  =    e_{i}  =   Q_{i};\hspace{1.5cm}		\sigma(y_{i})  =  
e_{i^{\star}} =  i P_{i}/\hbar  =  D_{i},\]
so that the algebra is generated by the set $ \{1, D_{i}, Q_{i}, E\} $ with
\be
	 [D_{i}, Q_{j}]  =  \delta_{ij};\quad 	E^{2}  = E;\quad 
D_{i}E  =  0;\quad EQ_{i}  =  0. \quad 	\forall i \in {\mathbb Z}
\ee
We see that this structure is a generalisation of the Heisenberg algebra.  

	The addition of this new element is not as arbitrary as may appear at first sight.  Hiley and Monk \cite{bhnm93} have shown how an element like $E$ arises naturally in the finite generalised Clifford algebra $ {\cal C}_{2}(n) (n \in {\bf Z}^{+})$.   This algebra is called the generalised Clifford algebra by Morris \cite{am67}, a quantum algebra which corresponds to a discrete phase space with periodic boundary conditions.  This polynomial algebra is generated over the complex field by the set
$\{1, U, V\}$ subject to
\[ UV  =  \omega VU;\hspace{1cm}		U^{n}  =  1;\hspace{1cm} 	V^{n}  =  1;\]
where   $\omega  =  \exp(2\pi i/n) $ and $n$ is an integer.

	This algebra is non-nilpotent and therefore one can construct at least one primitive idempotent from its elements \cite{bhpd}.  These elements again play an analogous role to the isotropic $r$-vector generated in the OCA and it is the presence of this idempotent that allows us to construct spinors in this generalised Weyl algebra.  
	It was shown in Hiley and Monk  \cite{bhnm93} (see also Weyl \cite{hw31}) that, in the limit $ n \rightarrow \infty $ when the phase
space becomes a continuum, this idempotent generalises to the $E$ introduced above. 

 When this element is omitted, the remaining structure is the usual Weyl algebra generated by the Heisenberg algebra.   What we suggest is that, as far as physics is concerned, the Heisenberg algebra \cite{bhnm93} generated by the elements $ \{1, P, Q\}$ is not the correct algebra required in quantum mechanics.  Rather it is the generalisation of this
algebra to including the idempotent $ E $ that is more appropriate.  
Indeed it is our claim that it is only this algebra that enables us to give a complete algebraic formulation of quantum mechanics as was discussed in Frescura and Hiley \cite{ffbh80}.

	We can now construct the symplectic spinor.   If, for simplicity we consider the case of $ r = 1,$ then this spinor can be written in the form
\begin{eqnarray}
		\psi  = \sum \lambda_{H} Q^{H} E	\label{eq:17}				
\end{eqnarray}
where the symbol $ \Phi^* $ has been replaced by the element $ E $.  Within the extended algebra, we can construct a minimum left ideal (\ref{eq:SMI}) as an exact analogue to the OCA \cite{ffbh80}, \cite{ibrt83}.

	We can also introduce a conjugate spinor which we define as a minimum right ideal. This can be written in the form
\be
		\phi  =  \sum \lambda_{K^*} E D^{K^*} 				
\ee
so that any element of the generalised SCA can then be written in the form
\be
		\Lambda  = \psi\phi  = \sum \lambda_{HK^*} Q^{H}E D^{K^*} .	\label{eq:19}		
\ee
The interesting feature of this combination of spinors is that the left spinor  only  involves position space operators, while the right spinor involves only momentum space operators.  In order to generate a minimum right ideal in terms of position operators we must introduce an element $ E^{\dagger} $,  the Hermitean conjugate of $ E.$ We will discuss this point further in section 5.2

\section{The Symplectic Spinor.}

\subsection{ The Physicist's Approach to the Symplectic Spinor}

	 In order to motivate the connection between the algebraic spinor and the  physicist's approach using the more familiar bra-ket notation, it is first necessary to recall the {\em standard} ket  introduced by Dirac \cite{pd39}.  To avoid a misunderstanding, we want to emphasise the the standard ket Dirac discusses in the 3rd edition of his classic book \cite{pd47} is not the usual ket, $|\psi\rangle$, used universally in quantum mechanics.  Dirac writes the standard ket as $\psi\rangle$, i.e. with the $|$ missing.  This is a seemingly  small notational change, but involves a big change in meaning.  We will now bring out this meaning and  show how it is related to  the element, $ E $, that we have introduced.

	Suppose $ \{A_{j}\}$ is some given complete set of dynamical operators, together with an eigenbasis  $ |a_{ij}\rangle. $  Two indices are needed here, one for a particular operator in the set and one for its set of eigenfunctions.  As is well known, we can associate a complex function with every ket $|P\rangle $ viz. 
$ \psi(a_{ij})  =  \langle a_{ij}|P\rangle$.
This means that in the basis $ \{|a_{ij}\rangle \},$ each function $\psi$ uniquely defines $|P\rangle $, so that we may write
\[|P\rangle  =  |\psi,\{A_{r}\}\rangle. \]
Suppose now that $ f(A_{r}) $ is any function of the observables $ \{A_{r}\}.$  Then $f(A_{r}) $ is itself an operator in its own right and we have
\[ \langle a_{j}|f(A_{j})|\Psi,\{A_{r}\}\rangle  =  f(a_{ij})\langle
a_{ij}|\Psi,\{A_{r}\}\rangle \]
				 \[  = f(a_{ij})\psi(a_{ij})  =  (f.\psi)(a_{ij}) \]
				 or
				 	     
\[f(A_{j})|\Psi,\{A_{r}\}\rangle  = |(f.\Psi), \{A_{r}\}\rangle
\hspace{1cm}\forall\hspace{0.5cm}  |\Psi,\{A_{r}\}\rangle.\]

 In particular, for $ \Psi = 1 $ we have
  \[     f(A_{j})|\{A_{r}\}\rangle  =  |f, \{A_{r}\}\rangle \]
Dirac argued that the vertical line $ | $ is not necessary once we have singled out the complete set  $\{A_{j}\} $ we are considering, so he then introduced the new notation by writing
\be
			       f(A_{j})|\hspace{0.2cm} \rangle =  f \rangle 
\ee	
He called $ \hspace{0.1cm} \rangle $ the standard ket for the complete set $\{A_{j}\} $.   

Notice that operating from the right of $\hspace{0.1cm}\rangle$ is not defined, so clearly the symbol defines a left ideal under left multiplication. Note also that different complete sets will require different standard kets, so that the symbol $\hspace{0.1cm}\rangle$ is playing the role similar to that of the idempotent $\epsilon$ in the OCA, but we are not in a position to identify the two approaches yet.  To do that we need to introduce a standard bra to define a right ideal.  Dirac does this by introducing the symbol $ \langle \hspace{0.1cm} $ so that every bra can be written as
  $ \langle \hspace{0.1cm}  \Phi(A_{r}) $.  We can write a general element in the form
\be
	\Lambda =  \Psi(Q)  \hspace{0.1cm}\rangle \langle
\hspace{0.1cm}  \Phi(D). 		\label{eq:genele}
\ee
 To see what this means, let us consider the polynomial algebra again and   write 
	\[\Psi(Q)  =  \lambda_{H}Q^{H} \hspace{1.5cm} 		{\rm and}  \hspace{1.5cm} 	\Phi(D)  = 
\lambda_{K^{\star}}D^{K^{\star}}, \]	
so that equation (\ref{eq:genele}) can be written in the form
\be
	\Lambda  =  \sum \lambda_{H}Q^{H}\hspace{0.1cm}\rangle \langle \hspace{0.1cm}
D^{K^{\star}}\lambda_{K^{\star}}.					 
\ee
Comparing this with eqation (\ref{eq:19}),  we see that $ E $ can be identified with the symbol  $ \hspace{0.1cm}\rangle \langle \hspace{0.1cm}  $.  Thus Dirac has effectively introduced an idempotent into the Heisenberg algebra.  Any element can be written in the general form
\be
\Lambda=\Psi_L(Q)\Phi_R(D)=\psi_L(Q)E\phi_R(D)		\label{eq:25}
\ee
where $\Psi_L(Q)=\psi_L(Q)E$ is an element of the left ideal and $\Phi_R(D)=\phi_R(D)E$ is an element of a right ideal.  Note the similarity between this equation and equation (\ref{eq:8}) for the OCA. 

Under the metaplectic transformation, the general element $ \Lambda $  transforms as
\be
	 \Lambda'=	M_{S}\Lambda M_{S}^{-1}  , 
\ee
where $ M_{S} $ is an element of the metaplectic group.  Analogous to the OCA we have $\Psi_L'=M_S\Psi_L$ and $\Phi_R'=\Phi M_S^{-1}$.  We will see later that it is the 'splitting' of the element $\Lambda$ that is responsive for the Gouy effect. 

It should be noted again that the right spinor is in momentum space and is not the hermitean conjugate to the left spinor.  As we will show, this asymmetry means that $ E $ is not hermitean so that  $ E^{\dagger} \neq  E.$  This in turn implies that the standard bra introduced by Dirac is not the standard bra we use above.  In fact while Dirac writes
 $\langle \hspace{0.2cm}     =  \hspace{0.2cm} \rangle^{\dagger},$     the bra we use above is  $ \langle \hspace{0.2cm}  \neq    \hspace{0.2cm} \rangle ^{\dagger}.$		
This asymmetry in the symplectic spinor is not present in the orthogonal spinor and has
 its roots in the notion of polarisation discussed in Guillemin and Sternberg \cite{gs77}.
 
 \subsection{The Metaplectic Covering in a Non-Commutative Phase Space}

Let us now consider two features that will make the language we are using clear.  Firstly let us recall how the bra-ket notation can be expressed in terms of matrices.  We write
\begin{eqnarray*}
|m\rangle\langle n|=E^{mn}=M\rangle\langle N= MEN
\end{eqnarray*}
where $M,N$ are the operators defined by
\begin{eqnarray}
M\rangle=|m\rangle\quad\quad N\rangle =|n\rangle   \label{eq:30}
\end{eqnarray}
Let us consider one of the terms in the sum (\ref{eq:29}) and write it as  
\be
		E^{mn}  =  (m!n!)^{-1/2} Q^{m}ED^{n}		\label{eq:31}	
\ee
and define $Q^{0}  =  1  =  D^{0}$,  so that $ E^{00}  =  E $.  Then it can be easily be shown
that
\be
		E^{mn}E^{ij}  =  \delta^{jm}E^{in}					
\ee 
thus the $ E^{mn} $  are linearly independent.  It can further be shown that 
\be
		1  =  \sum_{r}  E^{rr}			\label{eq:22}	
\ee
and
\be
	Q  =  \sum_{n} (n + 1)^{1/2} E_{n+1,n}, \hspace{1cm}   {\rm and}  \hspace{1cm} 	D  =  \sum_{n}
(n+1)^{1/2} E^{n,n+1}.
\ee
So all the generators of this algebra can be written in terms of the elements $ E^{mn} $   which therefore constitute a basis of the algebra.  In fact equation (\ref{eq:22}) is the defining relation for a matrix basis in a full matrix algebra.

\subsection{The Metaplectic Group in the Boson Algebra.}

Let us now express the $ Q, D $ in terms of a new set of elements which we write as  $ a, a^{\dagger} $. The transformation is well know, namely,
\be
a  =  (Q + D)/\sqrt{2} \quad{\rm and} \quad	a^{\dagger}   =   (Q - D)/\sqrt{2}.	\label{eq:35} \ee
We also introduce the idempotent operator $ V $ to play a role analogous to $ E $.  We then have the relations
\be
	[a , a^{\dagger}]_-  =  1,\quad	aV  =  0,\quad	Va^{\dagger}  = 
0,\quad V^{2}  =  V
\ee
 These relations define a set of boson annihilation and creation operators.  It should be noted that these elements are vectors because in general we write
\be
a^\mu  =  (Q^\mu + D^\mu)\sqrt{2} \quad	{\rm and} \quad	a^{\dagger\mu}   =   (Q^\mu - D^\mu)/\sqrt{2}.	 \ee 
It our example to illustrate how the formalism works, we have chosen $\mu=1$ for simplicity. With this change of variables equation (\ref{eq:31}) becomes
\be
		V^{mn}  =  (m!n!)^{-1/2} (a^{\dagger})^{m} V a^{n}	\label{eq:38}		
\ee
so that
\[a  = \sum_{n} (n+1)^{1/2} V^{n,n+1} \quad   {\rm	and}  \quad	a^{\dagger}=\sum_{n} (n + 1)^{1/2} V^{n+1,n} .\] 
  In this case it is easy to show that V is the projection to the vacuum state.  Thus if we write
\be
			V  =   |0\rangle\langle 0 |					
\ee
equation (\ref{eq:38}) then becomes
\be
V^{mn}  =  (m!n!)^{-1/2} (a^{\dagger})^{m}  |0\rangle \langle 0 | a^{n}.		
\ee

	We can now introduce the symplectic spinors
\be
	\Psi(a^{\dagger} )  =  c_{H}(a^{\dagger} )^{H}V \quad	{\rm and} \quad
\Phi(a)  =  c_{K}Va^{K}.   \label{eq:41}
\ee
where $ c_{H} $ and $ c_{K} \in \bm K.$
In this case the right spinor is the hermitian conjugate of the left spinor.  Hence
		\[(\Psi(a^{\dagger}  ) )^{\dagger}   =  \Psi^{\dagger} (a)  =  \Phi(a) \] 
Equation (\ref{eq:41}) then implies that $ V^{\dagger}   =  V.$
  
	Since we pass from the $ (Q, D)$ to the $(a, a^{\dagger})$ by the transformation (\ref{eq:35}), there
is a relation between $ E $ and $ V.$  In fact it was shown in Frescura and Hiley \cite{ffbh84} that
\be
		V  =  \lambda \exp[-Q^2/2] E \exp [D^2/2]		
\ee
and
\be
		E^{\dagger}  =  \Lambda \exp[-D^{2}/{2}] \exp[-Q^{2}/{2}]  E \exp[D^{2}/{2}] \exp[Q^{2}/{2}] 	 
\ee
Thus $ E $ is not Hermitean.  This property of the projector is not apparent if we use the 
standard bra-ket notation as used by Dirac.  The non Hermitean property of $ E $  also helps to understand why, if the symplectic spinor is only a function of $ Q $,  its conjugate-spinor is only a function $ D $.  If $ E $ had been Hermitian then spinor and co-spinor could have been functions of the same variables as they are in the Fock representation.

\section{Phase change occurring on going through a focal point: the Gouy Effect.}

\subsection{Calculation of the phase change using SCA.}

	We now have sufficient mathematical details to illustrate the use of the SCA in physics. The example we will give as an illustration is the Gouy effect which arises from the double cover of the symplectic group,namely,  the metaplectic group.  This effect has its roots in the properties of the symplectic spinor which provides an analogue of the orthogonal spinor.
	
	Let us briefly recall how  the physical consequences of the double covering of the rotation group is demonstrated experimentally.  Suppose we have a beam of neutrons incident on a beam splitter so as to produce two spatially separated beams of equal amplitude as shown in the Figure 1.
	
\begin{figure} [h]
\begin{center}
\includegraphics[width=4in]{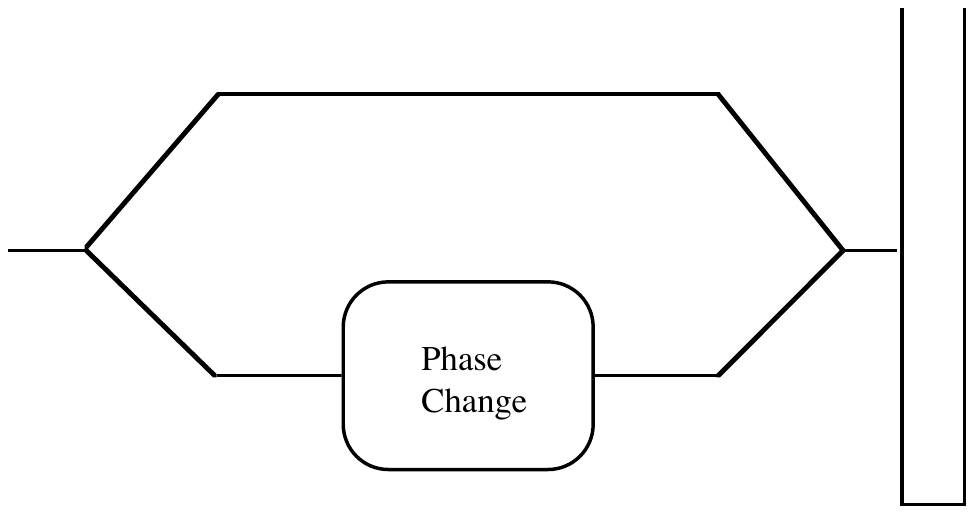}
\vspace{-4.6in}
\end{center}
\caption{ Phase change induced in one beam to produce a fringe shift.  The spinor must be rotated through $ 4\pi $ to return to the original pattern.}
\end{figure}
									
Let the lower  beam pass through a magnetic field designed to induce a phase change of $\theta$ relative to the upper beam (see Aharonov and Susskind \cite{yals67}).  Let the two beams subsequently  recombine so that the only phase difference is induced by the magnetic field.  The initial spinor at the first beam splitter is 
\[ |\Psi_{i} \rangle  =  |\psi_{1} \rangle   +  | \psi_{2} \rangle.\]
When the two beams are recombined, the final spinor is
\be
			|\Psi_{f} \rangle  = |\psi_{1} \rangle   +  g(\theta/2)|\psi_{2} \rangle .		
\ee
If  $ \theta  = 2\pi $,  the final spinor is

		  \[	|\Psi_{f}\rangle  =  |\psi_{1} \rangle   -  |\psi_{2} \rangle .\]

Clearly it is only after the second beam has been subjected to a phase change of $ 4\pi $ that the interference pattern is the exactly the same as if no phase change in the lower beam.  Experiments realising this effect have been performed by Rauch {\em et al} \cite{rwbf}, Werner {\em et al} \cite{wcoe}, Klein {\em et al} \cite{ko76} and Rauch {\em et al} \cite{rzbf} 
and they clearly demonstrate the physical consequences of the double cover.

Now let us turn to consider how the double cover of the symplectic group could be demonstrated in an experiment.  In order to illustrate how this works in a particular case, let us take an example from optics.  Suppose we consider the thin lens arrangement shown in figure 2. 
											
											
\begin{figure} [h]
\begin{center}
\includegraphics[width=4in]{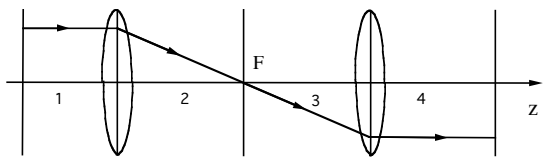}
\vspace{-4.4in}
\end{center}
\caption{ A pair of identical thin lenses that pass the ray through the focal point F.}
\end{figure}											
											
First let us consider how we trace  the ray through the lens system using the Gaussian optics, which, of course, means using the symplectic group.
 The incident ray is parameterised by the matrix $\left (\begin{array}{clcr}{q}\\ {p} \end{array}\right)$, $\;q $ being the vertical distance of the ray from the optic axis and $p$ is $ n\theta $ where $n$ is the refractive index of the medium and $ \theta $ is the angle the ray makes with the optic axis. 
 
  For  the case shown in Figure 2,  initially the ray is above and parallel to the axis so $ p = 0,$.   Thus the incident ray is described by  $
\left(\begin{array}{clcr}{q}\\ {0}\end{array}\right)$.   The progress of this ray through the first lens (regions 1 and 2) is described by the following set of three symplectic matrices
\be
		{\bm S}  = \left(\begin{array}{clcr} 1 & f\\ 0 & 1\end{array}\right) \left(
\begin{array}{clcr}1 & 0 \\ -1/f & 1 \end{array}\right) \left(\begin{array}{clcr} 1 & f\\ 0 &
1\end{array}\right)  =  \left(\begin{array}{clcr} 0 & f\\
 -1/f & 0 \end{array}\right) 
\ee
The ray that arrives at region 4 in Figure 2 is then found by applying this transformation  twice we obtain
\be
{\bm S}^{2} \left(\begin{array}{clcr} q \\ 0 \end{array}\right)    =  \left
(\begin{array}{clcr} -q \\ 0 \end{array}\right)  			
\ee
which shows that there has been an overall sign change which is exactly what we expect  from ray optics, namely, the final ray is below and parallel to the optic axis.  It is interesting to examine what happens in the $ (q,p) $ phase space plane as the ray passes through regions 1 to 4.  The evolution of the phase vector through the lens system is shown in the figure 3 and it can clearly be seen to rotate through an angle $ \pi $.

											
\begin{figure} [h]
\begin{center}
\includegraphics[width=4in]{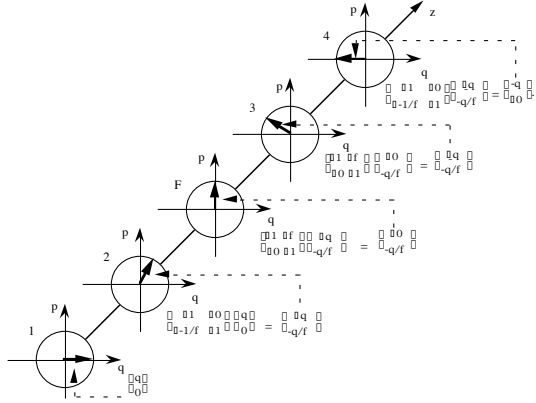}
\vspace{-3in}
\end{center}
\caption{Rotation of vector in phase space as the ray travels through the lens system shown in figure 2. }
\end{figure}															
	
		In order to return the ray to its original state we must add another two identical thin lenses to this set up.  This is shown in Figure 4.  We then apply the same transformation to complete tracing the ray until it reaches region 7.  Then it is easy to show that the final ray is again described by $ \left(\begin{array}{clcr} q\\ 0 \end{array}\right) $.  This is
consistent with the result $ \bm{S}^{4}  =  \bf{1} $. 


\begin{figure} [h]
\begin{center}
\includegraphics[width=4in]{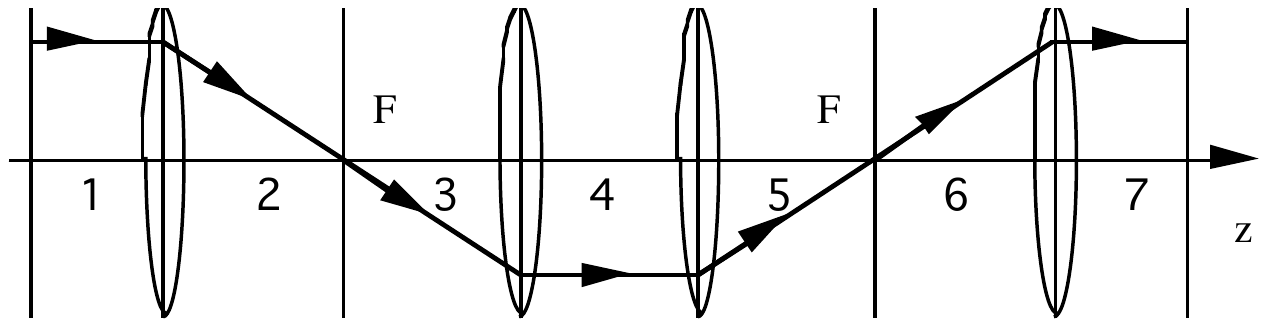}
\vspace{-5in}
\end{center}
\caption{Four identical thin lenses set-up. }
\end{figure}				
 
	In order to take account of the wave properties of the light, we must introduce the symplectic spinor and examine happens to the symplectic spinor under a metaplectic transformation. The metaplectic transformation corresponding to $ \bm{S} $ is
\be
		{\bm M}_{s} = \exp[-if P^{2}/2] \exp[- i Q^{2}/2f] \exp[- if
P^{2}/2] 	 
\ee
where we have used the correspondences
\be
\exp[-if P^{2}/2]  \leftrightarrow \left(\begin{array}{clcr} 1 & f\\ 0 &
1\end{array}\right)	\label{eq:pf}	\ee	
and
\be
\exp[-i Q^{2}/2f] \leftrightarrow \left(\begin{array}{clcr} 1 & 0\\ -1/f &
1 \end{array}\right) 		
\ee
These elements then act from the left on the general spinor 
\[ \Psi_L(Q)  =  \sum
\lambda_{H}Q^{H}E. \]

	In order to be more specific, the spinor associated with the ray above is $ \Psi_L(Q)  =  QE $.  Then we need to evaluate $ \bm{M}_{s} \Psi_L(Q)  =
\Psi^{\prime_L(Q)}. $  This is achieved by noting that
\be 
		{\bm M}_{s} Q E = \bm MQ\bm M_s^{-1} \;\bm M_sE=Q^{\prime} {\bm M}_{s} E,	\label{eq:45}
 \ee
where $ Q^{\prime}  =  {\bm M}_{s}Q{\bm M}_{s}^{-1}=\bm SQ $.  The remaining task is to evaluate the effect of the metaplectic transformation on $ E.$  This can be done in a straight forward manner.  For the ray arriving at region 4 in Figure 4 it is
\begin{eqnarray}
{\bm M}_{s}E  =  \exp[- if P^{2}/2] \exp[- i Q^2/2f]  \exp[- if
P^{2}/2] E \\ \nonumber=   \exp[- if P^{2}/2] \exp[- i Q^{2}/2f]  E,\hspace{2.2cm}
\label{eq:mps}
\end{eqnarray}
which must now be evaluated. An attempt to use the Baker-Hausdorff method quickly becomes unmanageable.  A much simpler way of proceeding is to
use the integral kernel method. (For full details see Moshinsky and Quesne \cite{nmcq71}).

The evaluation is simpler if we use the fact that  this phase change in the Dirac notation corresponds  to 
	\[	|\Psi^{\prime} \rangle =  {\bm M}_{s} |\Psi \rangle. \]
In the Schr\"{o}dinger representation this becomes
	\[	\langle x |\Psi^{\prime} \rangle  =  \Psi^{\prime} (x)  =\int_{- \infty}^{\infty}  
\langle x |\bm{M}_{s}|x^{\prime} \langle x^{\prime} | \Psi \rangle dx^{\prime}  =
\int_{-\infty}^{\infty}\bm{M}_{s}(x,x^{\prime})\Psi (x^{\prime})dx^{\prime}  \label{eq:wa} \] 
In this representation
\begin{eqnarray}
	\exp[- if P^{2}/2]     \leftrightarrow  \frac{1}{\sqrt{2\pi if}}\int_{-
\infty}^{\infty} \Psi(Q^{\prime})\exp\left[-\frac{(Q^{\prime} -Q)^{2}}{2if}\right ]dQ^{\prime} 
\end{eqnarray}
 and
\be
	\exp[-i Q^{2}/2f] \;	  \leftrightarrow	\;	\exp[- iQ^{2}/2f] \Psi(Q^{\prime}).		\label{eq:54}
\ee
These expression are derived in Guillemin and Sternberg \cite{vgss84}.  Using these expressions we can evaluate the expression in equation (\ref{eq:mps}).  We find
\begin{eqnarray}
{\bm M}_{s}^{2} E  =  \exp[- ifP^{2}/2] \exp[- i Q^{2}/2f] \exp[-
if P^{2}/2] {\bm M}_{s}E  \\\nonumber =  (-i) E  =  \exp(-i\pi/2) E.\hspace{4cm}
\end{eqnarray}
	This shows that this particular metaplectic transformation on E causes the spinor to undergo a phase transformation of $ \pi/2.$  From equation (\ref{eq:45}) we find that the spinor, $ \Psi_L(Q)  =  QE $, becomes
\[	{\bm M}_{s}^{2}QE  =  Q^{\prime} {\bm M}_{s}^{2}E  =  \exp(i\pi)Q\exp(-i\pi/2)E  = \exp(i\pi/2)QE \]
This result shows a $\pi/2 $ phase change for the ray ending in region 4 of Figure 4.  This agrees with the result derived from wave theory for a cylindrical wave. (See Sommerfeld \cite{as54}).  If we evaluate $ {\bm M}_{s}^{4} $ we find a phase change of $ \pi $ so that we have to pass the through a further set of four lenses in order to return the ray to its original phase.  We have calculated this result using the SCA.  In the next section we show how  the same result is calculated using wave theory.
  
\subsection{Calculation of phase change using Wave Theory}

	It is not immediately clear that the `phase factor' obtained here is the same as the phase
 factor found in wave theory.  To show that this is so, let us briefly recall how this phenomena is dealt with in conventional paraxial wave theory.  In this approximation the wave equation is written in the form
	\[ \frac{\partial^{2}R}{\partial x^{2}}  +  \frac{\partial^{2}R}{\partial y^{2}}   +  2ik\frac{\partial R}{\partial z}  =  0, \]
which is just a two dimensional Schr\"{o}dinger equation of a particle of mass $ k$ with $\hbar = 1$. The role of time is now played by $z$.  The general solution of this equation is
		\[\psi({\bm r},{\bm r}_{0})  = \int_{-\infty}^{\infty} \frac{\exp[-ik
\rho ({\bm r},{\bm r}_{0}]}{\rho ({\bm r},{\bm r}_{0})}   d {\bm r}, \]
where
	\[ \rho ({\bm r},{\bm r}_{0})  =  \sqrt{(x - x_{0})^{2}+(y - y_{0})^{2}+(z
-z_{0})^{2}}. \]
If we want the paraxial cylinderical wave solution, we expand the square root so that the wave function  can be written in the form 
	\[ \psi ({\bm r}, {\bm r}_{0})  \approx \frac{1}{z - z_{0}}  \exp[- ik(z -
z_{0}]\int_{-\infty}^{ \infty} \exp\left[- ik \frac{(x - x_{/0})^{2} + (y-y_{0})^{2}}{2(z -
z_{0})}\right] dxdy \]
	This wave does not fall off with distance so, following Siegman \cite{as86}, one adds a complex quantity to $ z - z_{0} $ so that
	\[	\xi (z)  =  z - z_{0} - \xi_{0} \quad	{\rm with} \quad  \xi (z)  \in
\mathbb C. \]
 One then writes
	\[ \frac{1}{\xi (z)}  =  \frac{1}{ \xi_{1}(z)}  - \frac{i}{ \xi_{2}(z)}  \quad  	
{\rm with} \quad  \xi_{1}(z), \;\xi_{2}(z) \in \mathbb R. \]
Then the wave function becomes
\begin{eqnarray}
\Psi ({\bf r},{\bf r}_{0})  =  \sigma \exp(i\theta)\exp[- ik(z - z_{0})]\hspace{4cm}\nonumber\\
\times \int_{-\infty}^{\infty} 
\exp\left[- ik \frac{(x - x_{0})^{2}}{2\xi_{1}(z)}\right]\exp\left[ - k\frac{(x - x_{0})^{2}}{2\xi_{2}(z)}\right] dx		\label{eq:52}	
\end{eqnarray}
where we have written  $ 1/\xi(z) = \sigma \exp(i\theta) $ and where $ \arctan \theta =  \xi_{1}/ \xi_{2} $ .  Here the angle $ \theta $ is the Gouy phase.
	
	In order to relate this to the algebraic approach we must first note that the matrix $  \left(\begin{array}{clcr} 1 & f\\ 0 & 1\end{array}\right) $is a special case of the general translation matrix $ 
 \left(\begin{array}{clcr} 1 & \zeta/n\\ 0 & 1\end{array}\right)$ 
where $ \zeta  =  z - z_{0} $ and $ n $ is the refractive index which, for convenience, we will take to be unity.  Then equation (\ref{eq:54}) can be written as
\begin{eqnarray}
\exp[- ( \frac{i\zeta}{2}) P^{2}] \hspace{0.3cm} \leftrightarrow \hspace{0.3cm}
\frac{1}{\sqrt{2\pi i(z - z_{0})}} \int_{-\infty}^{\infty}\Psi(Q^{\prime})\exp\left[-
\frac{(Q^{\prime} -Q)^{2}}{2i\zeta }\right]dQ^{\prime}	\label{eq:55}
\end{eqnarray}
Comparing this with equation (\ref{eq:52}), we see that if we identify $ Q^{\prime} $ with $ x $ and $ Q $ with $ x_{0} $, we establish the close similarity between the two approaches.  It is then possible to relate the second Gaussian term with the spinor  $ \Psi (Q^{\prime} - Q).$   Thus we can express this relationship in an algebraic form where the symplectic spinor
in this case can be written in the form
		\[ \Psi  =  \exp\left[ - k\frac{(Q - Q_{0})^{2}}{2\xi_{2}(z)}\right] E. \]
The propagator component of $\bm M_s$ can  then be written as
		 \[  \exp\left[- \frac{if}{2} P^{2}\right] \exp\left[ - k \frac{(Q - Q_{0})^{2}}{2\xi_{2}(z)}\right] E \]

	As we saw in equation (\ref{eq:52}), the result of this particular transformation when $ \xi_{2}
	\rightarrow 0,$ produces a phase change $\pi/2 $  .  Comparing this result with equation (\ref{eq:55}) enables us to identify this phase with the Gouy phase.
	
\subsection{Experimental Demonstration of Gouy Effect.}

Finally we recall how it is possible to demonstrate the effect of this phase transformation in an experimental situation.  Let us return to the beam splitting situation discussed in equation (\ref{eq:45}) above. Call the beam reflected off the plane mirror beam 1 and the beam reflected from the concave mirror beam 2.   When these beams overlap an interference fringe patter is produced.  We then observe the change in this interference pattern 	
	(a) before beam 2 passes through a focus, and then  
	(b) after beam 2 passes through a focus. 
	
There is a discontinuous shift in the interference fringes as one passes through the focal point.  This is a direct result of the phase change occurring when a beam passes through a focal point.  Indeed this is precisely what was demonstrated by Gouy in 1890. (See Figure 5).  An extremely elegant way of demonstrating the effect for microwaves will be found in Carpenter \cite{cc59}.
										
\begin{figure} [h]
\begin{center}
\includegraphics[width=4in]{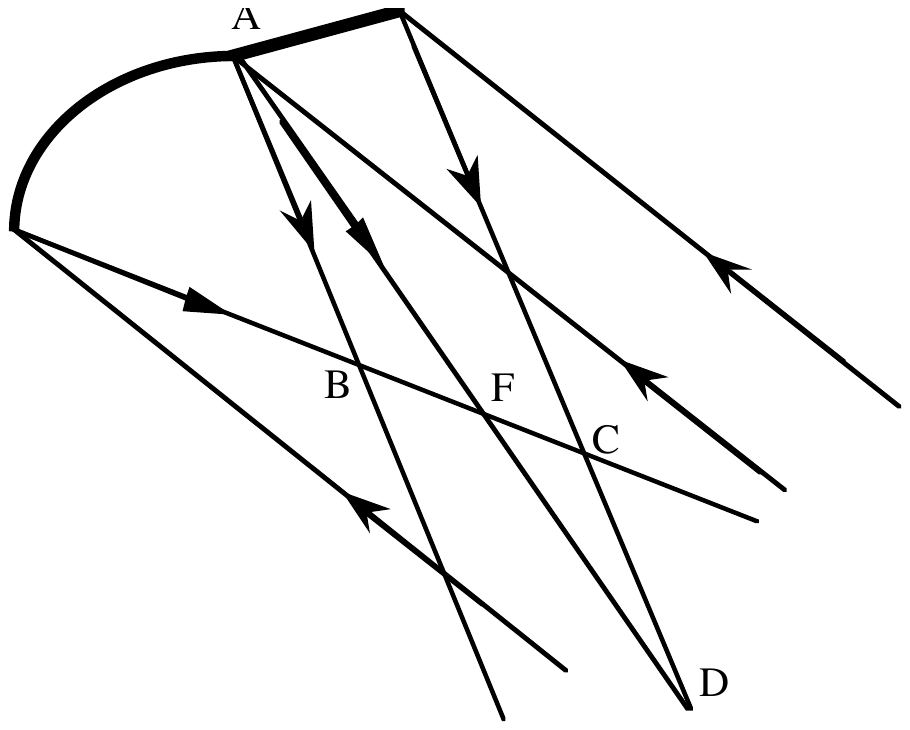}
\vspace{-4.2in}
\end{center}
\caption{The Gouy experiment with cylinderical mirror.  The interference fringes in region ABF are $ \pi /2 $ out of phase with those in region FCD.}
\end{figure}

\subsection{The relation of the Gouy phase to the Berry Phase}

	In the previous section we have shown the relation between the Gouy phase and the metaplectic group.   Our algebraic approach, emphasising the role played by the SCA  and the symplectic Clifford group is new, although the role of the metaplectic group has been pointed out by a number of authors, among them are Bacry and Cadilhac \cite{bc81} and more recently Simon and Mukunda \cite{ rsnm93}.  These authors have  also shown how the Gouy phase can be related to the Berry phase.  It is easy to see this connection in different way by using the optics example described in the previous section.  
	
	To bring out this connection we will use the ideas developed in Aharonov and Anandan \cite{yaja87}.  To this end we construct a fibre bundle with the $ (q, p) $-phase space, $ W,$ acting as  the base manifold, the symplectic spinor space, $ S_{K} $,  as the fibre and the metaplectic group acting as the structure group.  The bundle has a projection $ \pi $, such
that
		\[	\pi:  W \times S_{K}  \rightarrow  W  \] 
The optical path through the lens system shown in figure 4 then corresponds to a circuit shown in figure 6.

										
\begin{figure}[h]
\begin{center}
\includegraphics[width=4in]{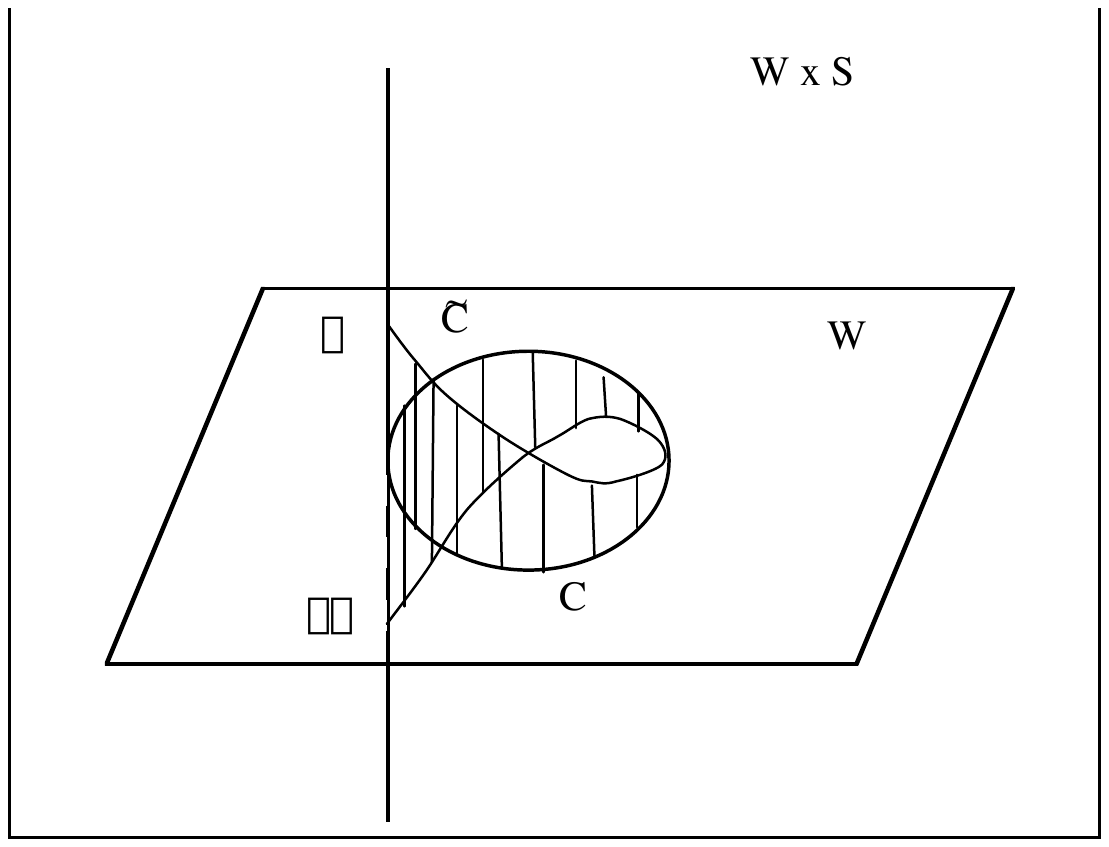}
\vspace{-4.1in}
\end{center}
\caption{Illustration of sign change in symplectic spinor after a $ 2\pi $ rotation in 	
	phase space.}
\end{figure}

 Consider the initial vector $ \in W, $ where $ w =\left(\begin{array}{clcr} q \\
0\end{array}\right) $.  As the light ray passes through the lens system, it undergoes a symplectic transformation $ \bm S^{4},$ returning it to its original position.  Since this completes the circuit, we can write $ \bm S^{4} = \bm S_{2\pi} $ .  Then under $ \bm S_{2\pi} $ we have
			\[ \bm S_{2\pi} : (w,\Psi) \rightarrow  (\bm S_{2\pi}w, {\bf M}_{\bm S_{2\pi}} \Psi) \]
where $ \Psi \in \bm S_{K} $ .  Now $ \bm S_{2p}  =  1 $ and $ {\bm M}_{\bm S_{2\pi}}  =  -1 $ so that

		\[	\bm S_{2\pi} : (w, \Psi) \rightarrow  (w, -\Psi).\]

	Thus we have to traverse the circuit again in order to return the symplectic spinor to its  original value.  In this way we have a direct analogy between the symplectic spinor and the orthogonal spinor.  For the orthogonal spinor the $ 2\pi $ rotation is in space while for the symplectic spinor it is in phase space.
	
	Since these results are quite general, one would expect other examples of the double covering of the symplectic group.  Again since our example exploits wave properties, we would expect similar effects as those found in Gaussian optics to be present in electron and atomic beam interference experiments.  To our knowledge no such effects have been investigated.

\section{Conclusion}

	We have seen how by adding a new primitive idempotent to the Heisenberg algebra as normally understood, we are able to define a symplectic spinor.  This brings out the striking analogy with the spinor used for spin-half particles.  This enables us to bring together two apparently different structures into different aspects of the same general general mathematical structure, namely, the Clifford algebra.  In fact this algebra is a way to study the covering spaces of the Lie groups involved.  This enables us to understand geometric phase in a new and unified way.  By exploiting this analogy we are able to see how the Gouy effect can be explained as a consequence of the double cover of the symplectic group.

\section{Acknowledgements}

	M. Fernandes wishes to thank the Brazilian National Research Council (CAPES)  for their support while this work was being completed.  We would also like to thank Professor F. A. M. Frescura for many helpful discussions.


\end{document}